\documentclass[11pt,twoside]{article}
\usepackage{asp2006}
\usepackage{graphics}
\usepackage{graphicx}
\usepackage{lscape}
\def\edcomment#1{\iffalse\marginpar{\raggedright\sl#1\/}\else\relax\fi}
\reversemarginpar

\begin{document}
\title{The Fabra-ROA Baker-Nunn Camera at Observatori Astron\`omic del Montsec: a wide-field
imaging facility for exoplanet transit detection}

\author{O. Fors\altaffilmark{1,2}, 
J. N\'u\~nez\altaffilmark{1,2}, 
J. L. Mui\~nos\altaffilmark{3},
F. J. Montojo\altaffilmark{3},
R. Baena\altaffilmark{2},
M. Merino\altaffilmark{1,2},
R. Morcillo\altaffilmark{3}
and
V. Blanco\altaffilmark{3}
}
\altaffiltext{1}{Observatori Fabra, Reial Acad\`emia de Ci\`encies i Arts de Barcelona, E-08002 Barcelona, Spain}
\altaffiltext{2}{Departament d'Astronomia i Meteorologia and Institut de Ci\`encies del Cosmos (ICC), Universitat de Barcelona (UB/IEEC), E-08028 Barcelona, Spain}
\altaffiltext{3}{Real Instituto y Observatorio de la Armada, E-11110 San Fernando, Spain}

\begin{abstract}

A number of Baker-Nunn Camera (BNC) were manufactured by Smithsonian Institution
during the 60s as optical tracking systems for artificial satellites with
optimal optical and mechanical specifications. One of them was installed at the
Real Instituto y Observatorio de la Armada (ROA).

We have conducted a profound refurbishment project of the telescope to be installed at Observatori Astron\`omic
del Montsec (OAdM)~\citep{Fors2009}. As a result, the BNC offers the largest combination of a huge FOV
(4.4$\deg$x4.4$\deg$) and aperture (leading to a limiting magnitude of V$\sim$20). 

These specifications, together with their remote and robotic natures, allows
this instrument to face an observational program of exoplanets detection by means of
transit technique with high signal-to-noise ratio in the appropiate magnitude range.  
\end{abstract}


\vspace{-1.0cm}
\section*{Refurbishment project}

The BNC was designed as a f/1 0.5m photographic wide field (5$\deg$x30$\deg$) telescope with a spot size smaller
than 20$\mu$m throughout the FOV.	

Among some others, the BNC has been refurbished following these phases: mechanical modification of the mount into
equatorial and motorization of the two axes, optical refiguring of the originally photographic curved FOV to
enable the use of a 4kx4k 9-$\mu$m custom-designed FLI ProLine CCD camera (see Fig.~\ref{fig:ccd}) and to comply
the Baker's original design spot size diagram, manufacture of a tip-tilt adjustable spider vanes assembly and
athermal CCD focus system (see Fig.~\ref{fig:spider}), mirror realuminization and outermost 50cm lens repolishing
to increase the throughput of the system, construction of a reinforced glass-fiber enclosure with sliding roof
which will host the BNC at OAdM (see Fig.~\ref{fig:enclosure}), development of an XML-based messaging protocol and
Java GUIs software, named Instrument-Neutral Distributed Interface (INDI), to control every device of the
observatory and schedule its operation both in remote and robotic modes~\citep{Downey2009}.

On 23 Sep 2008, the BNC successfully saw first technical light at ROA testing site (see Fig.~\ref{fig:m31}). Note
this image was still taken with unpolished 50cm lens and non-realuminizated mirror. The definitive commissioning
at OAdM is expected by early spring 2010.

\setcounter{figure}{0}
\begin{figure}[!h]
\begin{minipage}[t]{0.5\linewidth}
\centering
\includegraphics[scale=0.56]{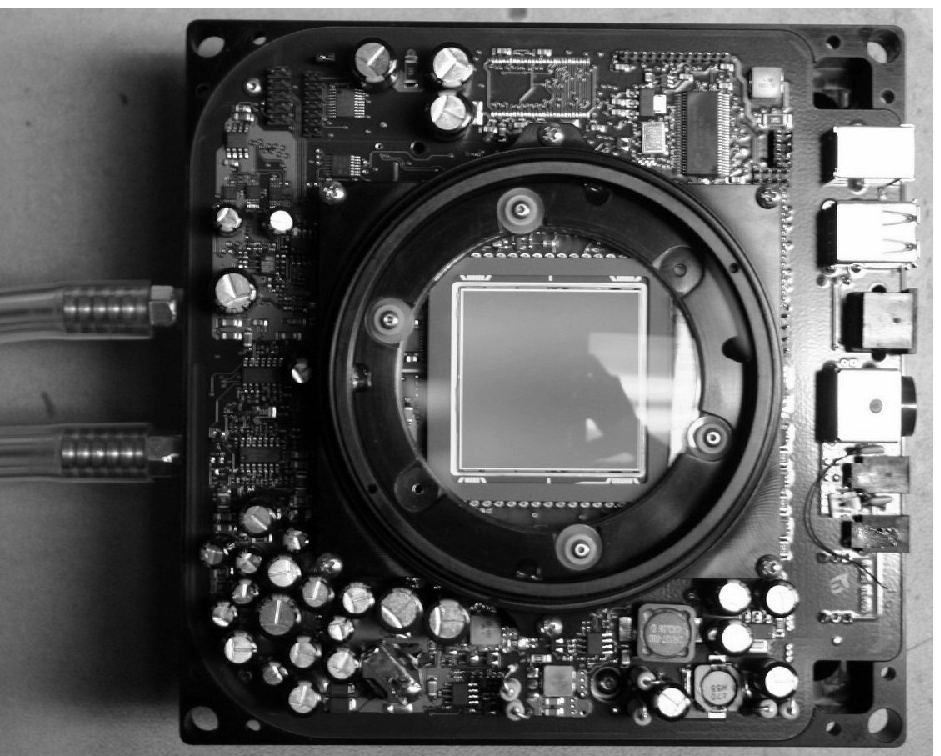}
\caption{Custom-designed FLI CCD with field flattener.}
\label{fig:ccd}
\end{minipage}
\hspace{0.1cm}
\begin{minipage}[t]{0.5\linewidth}
\centering
\includegraphics[scale=0.71]{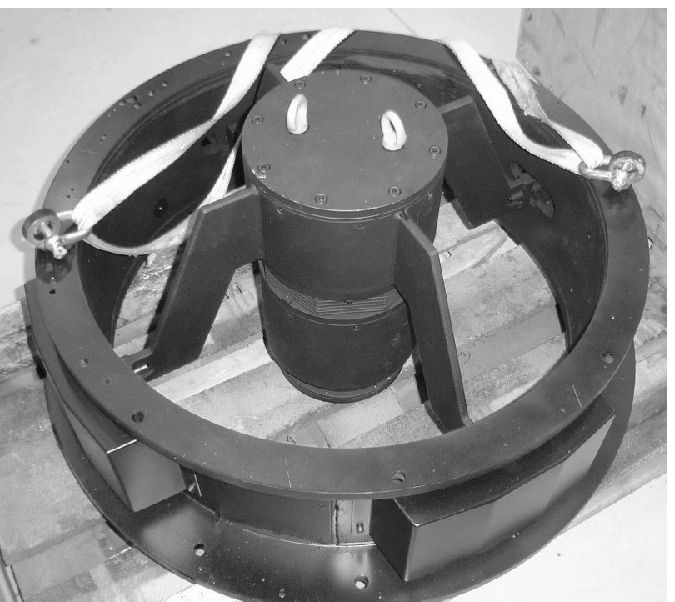}
\caption{Spider vanes assembly and focus system for the CCD.}
\label{fig:spider}
\end{minipage}
\end{figure}

\begin{figure}[!h]
\begin{minipage}[t]{0.5\linewidth}
\centering
\includegraphics[scale=0.58]{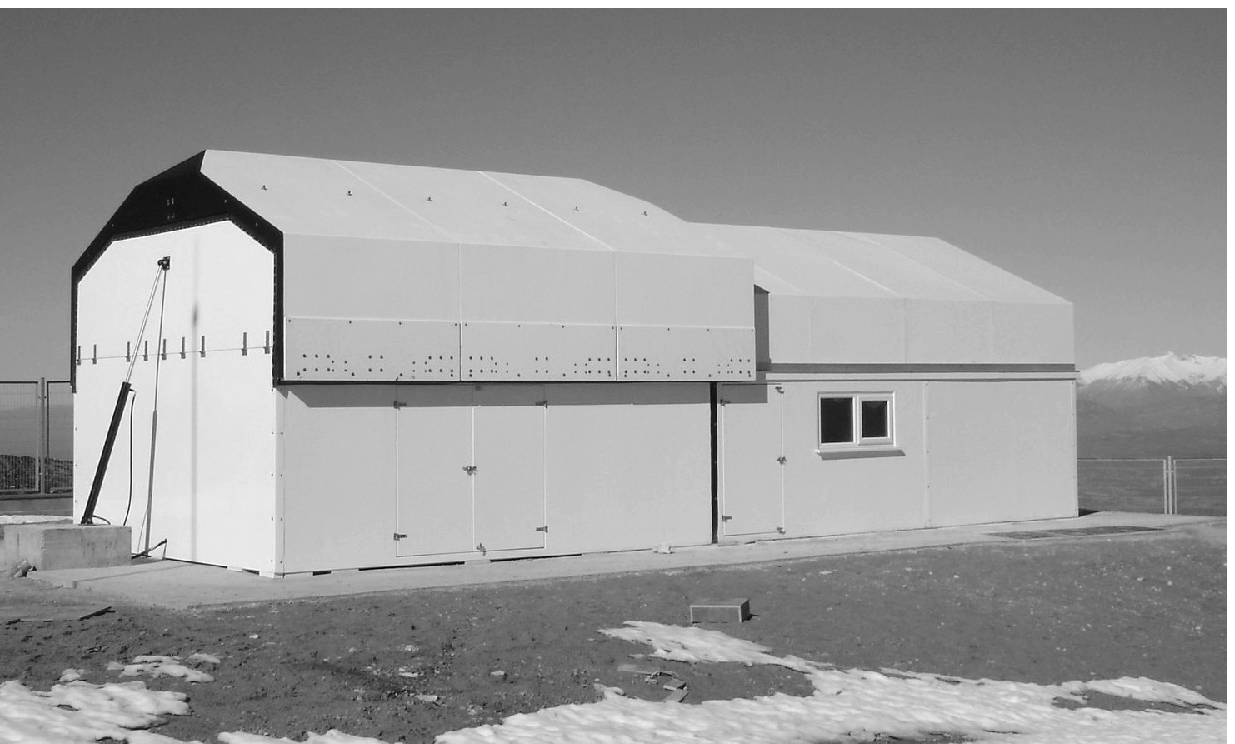}
\caption{Reinforced glass-fiber enclosure at OAdM.}
\label{fig:enclosure}
\end{minipage}
\hspace{0.1cm}
\begin{minipage}[t]{0.5\linewidth}
\centering
\includegraphics[scale=1.05]{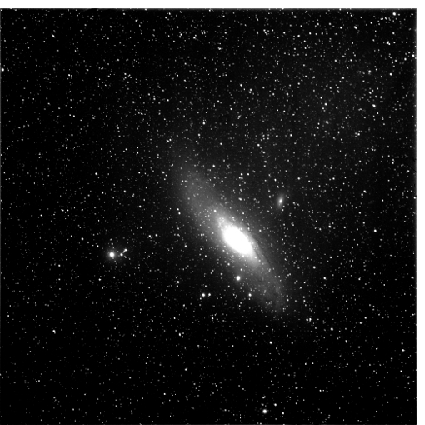}
\caption{First technical light of M31 at ROA on 23 Sep 2008.}
\label{fig:m31}
\end{minipage}
\end{figure}

\section*{Transit exoplanet detection}
The robotic nature of the BNC, its huge FOV and its considerable aperture, enables the telescope to succesfully
detect transits of exoplanets. This expectation is supported by the fact that the Automatic Patrol Telescope
(APT), originally a BNC twin of ours and, which after a similar refurfishment (although with an smaller FOV and
less sensitive CCD), has succesfully compiled the UNSW Extrasolar Planet Search 2004-2007 catalogue of exoplanets
candidates~\citep{2008MNRAS.385.1749C}. This catalogue shows that BNC-based cameras can
accomplish millimagnitude photometry at least up to V$\sim$14 magnitude.

\acknowledgements
This work was supported by the Ministerio de Ciencia e Innovaci\'on of
Spain 
and by Departament d'Universitats, Recerca i Societat de la Informaci\'o of the
Catalan Goverment.

\end{document}